\documentclass{aastex}
\begin{document}
\newcommand\kms{~km~s$^{-1}$}
\shorttitle{SMC Wolf-Rayet Stars}
\title{A Search for Wolf-Rayet Stars in the Small Magellanic Cloud}

\author{Philip Massey\altaffilmark{1} }
\affil{Lowell Observatory}
\affil{1400 W. Mars Hill Road, Flagstaff, AZ 86001}
\email{massey@lowell.edu}

\and

\author{Alaine S. Duffy\altaffilmark{1,2}}
\affil{Gettysburg College}
\affil{Box 756, Gettysburg, PA 17325}
\email{s432014@gettysburg.edu}

\altaffiltext{1} {Visiting Astronomer, Cerro Tololo Inter-American
Observatory,  National Optical Astronomy Observatory (NOAO), which is operated by the Association of Universities for Research in Astronomy, Inc. (AURA) under
cooperative agreement with the National Science Foundation (NSF).}

\altaffiltext{2}{Participant in the NSF's Research Experiences for
Undergraduates program, Summer 2000.}

\begin{abstract}
We report on a comprehensive search for Wolf-Rayet (W-R) stars in
the SMC using interference-filter imaging.  Photometry of over 1.6 million
stellar images on multiple, overlapping fields covering 9.6 square-degrees
found the previously known W-Rs at very high significance levels, two known
Of-type stars, plus additional candidates which we examined with slit
spectroscopy.  We discovered two new Wolf-Rayet stars, both
of type ``WN3+abs", bringing the total number in the SMC to 11.
We discuss their spectra, as well as reclassifying the previously known ones
with our new data.  Our survey also revealed 4 newly found Of-type
stars, including one of O5f?p type, which is 
one of the earliest-type stars known
in the SMC.  Another newly identified Of star is AV~398 (O8.5~If), a star
often used in extinction studies under the assumption that it is of early B type.  We recover S18 (AV154), a B[e] star whose spectrum currently lacks
He~II $\lambda 4686$ emission, but which must have had strong emission a
year earlier; we compare this star to S~Dor, suggesting it is indeed a
luminous blue variable (LBV).  We also find a previously unknown
symbiotic star.  whose spectrum
is nearly identical to the Galactic symbiotic AG~Dra.  More important, perhaps, than
any of these discoveries is the demonstration that there is not
a significant number of W-Rs waiting to be discovered in the SMC. The number
of W-Rs is a factor of 3 times lower in the SMC (per unit luminosity) than in
the LMC.  This strongly suggests that at the low metallicity which
characterizes the SMC only the most massive stars can evolve to W-R type.
\end{abstract}

\keywords{Magellanic Clouds --- stars: early-type --- stars: evolution ---
stars: emission-line, Be --- stars: Wolf-Rayet}

\section{Introduction}

For many years, our knowledge of the Wolf-Rayet (W-R) content of the SMC
was thought to be complete, with a total of eight discovered. Of these,
four had been found by general studies of the SMC's blue stellar
population (see Breysacher \& Westerlund 1978), with an
additional four found by an objective prism survey
targeted specifically at detecting W-Rs by using an interference filter
to reduce sky background and crowding (Azzopardi \& Breysacher 1979).
There matters stood until the discovery of Morgan, Vassiliadis, \& Dopita
(1991) of a ninth W-R. This star was well within the boundary surveyed by
Azzopardi \& Breysacher (1979), but a few tenths of a magnitude fainter
than any of the others.

Wolf-Rayet stars are the evolved, He-burning descendents of the most
massive stars, and their strong emission lines allow them to serve as
important tracers of the young stellar population in other galaxies
(Massey 1985).
The number of 
W-R stars in the SMC is of interest, as it appears to be
disproportionately low compared to that in the
LMC, where roughly 130 are known (Breysacher, Azzopardi, \& Testor 1999).
The (visual) luminosity of the SMC is 
only about a factor of 3.6 less than that
of the LMC ($M_V=-17.1$ and $M_V=-18.5$ respectively, van den Bergh 2000a), and, all other things being equal, we would thus expect $\sim$35 W-Rs in the SMC. Thus the presence of only 9
W-Rs suggests an under-abundance by a factor of $\sim 3.5$.
  
This scarcity, if real, is telling us
something interesting either 
about the progenitor massive star population in the SMC,
or about how its massive stars evolve in the SMC.  Let us examine these two
possibilities in turn. 

Since the visual luminosity
is dominated by an older population,  it could simply
be that the massive stars from which W-R evolve are scarcer in the SMC due to
a steeper initial mass function (IMF) and/or a lower current star 
formation rate (SFR).  What is the available evidence?
Massey (1998)
finds that 
the IMF
for massive stars appears to be indistinguishable between the SMC and the
LMC (and with the Milky Way for that matter).  In that case, a lack
of massive progenitors would require a
lower SFR in the SMC.
Could the current
SFR be 4 times lower in the SMC than in the LMC?  Let us compare the number
of OB associations in the two Clouds.  Hodge (1985) catalogs 70 regions
in the SMC, compared with the 120 found by Lucke \& Hodge (1970) in the LMC.
(Care was taken to use the same selection criteria in the
two galaxies, and indeed the size distributions are indistinguishable.)
This would suggest that the SMC has a SFR a factor of 2 {\it higher} per
unit luminosity than does the LMC. A better measure of the
SFR woud be the
integrated H$\alpha$ luminosity.  However, what is available is 
not quite this, but rather the total of all the H$\alpha$
emission from the discrete H~II regions.  Kennicutt (1991) finds that
these require an ionizing flux of $5 \times 10^{51}$~photons~s$^{-1}$ for the SMC, and 
$3\times 10^{52}$~photons~s$^{-1}$ for the LMC. The factor of 6 difference would suggest that the SFR
is 1.6 times lower per unit luminosity in the SMC. These totals do not include losses from density-bounded H~II regions, which should result
in diffuse
emission; such a correction may or may not be the same for the two Clouds. In any event, these data suggest that a lack of massive stars is not the primary cause for the factor of 3--4
under-abundance of W-Rs in the SMC compared to the LMC. 

On the other hand, we have good reason to suspect that the low metallicity of
the SMC will have a significant effect on the evolution of massive stars, as
there are smaller mass-loss rates at a given luminosity. In the
``Conti scenario" (Conti 1976; Maeder \& Conti 1994) Wolf-Rayet stars are
produced as a massive star loses mass via stellar winds; during the He-burning
stage the outer layers have been sufficiently stripped to reveal the H-burning
products at the surface and the star is spectroscopically identified as
a WN-type Wolf-Rayet.  Further mass-loss leads to a WC-type, with the products
of He-burning (C, O) revealed in the spectrum. The strong, broad emission
lines are the result of a highly extended and rapidly expanding stellar
atmosphere; e.g., optical depth unity occurs at a point in the stellar wind.
We would thus expect that stars
that 
evolve to  W-Rs in the SMC would have to be of higher luminosity 
(and mass) than those in galaxies with higher metallicity. 
(The oxygen abundance
is a factor of $\sim$2--3$\times$ lower 
in the SMC than in the LMC
(Lequeux et al.\ 1979; Russell \& Dopita 1992).
Indeed, the high proportion of WN-type Wolf-Rayets to those of WC-type in 
the SMC (8:1) is  consistent with the
hypothesis that WCs are harder to produce at low metallicities due to smaller
mass-loss rates, and
is in accord with the SMC's metallicity compared to
other Local Group galaxies (Massey \& Johnson 1998).  Recent studies
of coeval clusters and associations suggest that ``cut-off" mass for
becoming a W-R star is about 70$\cal M_\odot$ in the SMC, while stars
with initial masses greater than 30 $\cal M_\odot$ become W-Rs in the LMC
(Massey,
Waterhouse, \& DeGioia-Eastwood 2000); in the Milky Way the limit may be as
low as 18$\cal M_\odot$ (Massey, DeGioia-Eastwood,
\& Waterhouse 2001).

However, the discovery of the ninth SMC W-R by Morgan et al.\ (1991) raises
the question of the completness of previous surveys.  We 
were naturally curious if a significant additional population of W-Rs
remained to be found in the SMC.  We have had good success in searching for
W-Rs in nearby galaxies using a set of interference filters optimized to
detect W-Rs (e.g., Armandroff \& Massey 1985, Massey, Armandroff, \& Conti
1986; Massey \& Johnson 1998).   Accordingly we decided to survey the
SMC using the CTIO/Michigan Schmidt with these filters during a gap in
our main observing program last year.  Follow-up spectroscopy this year
identified 8 hitherto unknown interesting objects in the SMC, 
including 4 ``Of" stars, a luminous
blue variable (LBV) candidate, a symbiotic star, and two newly found
Wolf-Rayet stars.  We describe our survey,  these newly found objects, and
the implications here.

\section{The Search for New Wolf-Rayet Stars}
\subsection{The Survey}
Images were taken with the CTIO 
Curtis Schmidt telescope during (UT) 1999 October 23-28.  The field-of-view
was $1.3^\circ \times 1.3^\circ$ using the Tektronix 2048 detector;
the scale was 2.3 arcsec pixel$^{-1}$.  Outside the central $1000 \times 1000$~pixel region
our 2-inch filters caused  vignetting, with the amount of light
decreasing to 30--40\% on the sides, and to 5--15\% in the extreme corners.
We covered a 9.6 square-degree area centered on
the SMC with 12 (very) overlapping  fields to make the effects
of this vignetting minor.  The region surveyed is shown in Fig.~\ref{fig:smc}, where
we have also indicated the area covered by Azzopardi \& Breysacher (1979).

Our interference filter set is described in detail by Massey \& Johnson (1998):
the {\it WC} filter is centered on C~III~$\lambda4650$, the {\it WN} filter is
centered on He~II~$\lambda4686$, and the continuum {\it CT} filter is
centered at 4750~\AA.  The band-passes are approximately 50\AA\ wide.
The effective wavelengths of these filters used in the fast
f/3.5 beam of the Schmidt were close to these nominal values 
(4646~\AA\ for the {\it WC} filter and 4685~\AA\ for the {\it WN} filter).

For each of the 12 fields, we obtained 3 consecutive 600 sec
exposures in each of
the 3 filters (i.e., 9 per field). The data were biased and trimmed in
the usual manner, with twilight exposures serving as our flat-fields.
Stars brighter than $V\sim11.5$ were saturated.

Owing to the under-sampling, we performed only aperture photometry on the
images.  Nevertheless, analyzing the frames was a daunting task, with
{\it over 1.6 million} stellar images measured.  Each image was treated
separately, with the photometry averaged for the 3 exposures through each
filter at the end.  A 2.5 pixel (5.8 arcsec) radius aperture was used throughout.  Observations of several galactic W-R stars were used to set
the instrumental zero-point of the {\it CT} magnitudes to that of  
$v$ (an emission-free equivalent of
{\it V} introduced by Smith 1968b) with the
assumption (borne out by the results) that there were minimal color effects.
Photometry from
field to field was remarkably consistent, with typical differences
$\sim0.01$~mag for bright stars from night to night.

In order to identify W-R candidates we obtained the magnitude differences
{\it WC--CT} and {\it WN--CT}
 for all the objects in a field.  Stars significantly brighter
in one of the emission-line filters than in the continuum filter were judged
to be likely candidates.  As in our earlier work, ``significantly brighter"
was determined by using the actual statistical uncertainties of the photometry;
a magnitude difference had to be at least five times this value 
(i.e., 5$\sigma$) to be
considered a viable candidate.  In addition, we ignored stars
which appeared only 0.05~mag brighter in one of the on-line filters. We expect
a difference of 0.25~mag to correspond roughly to an emission-line
equivalent width (EW) of 
$-10$~\AA, about the limit of Wolf-Rayet stars (Massey \& Johnson 1998 and
discussion therein).  Our photometry went sufficiently deep that such a star
should be readily detected at $V=15.5$, corresponding to $M_V=-3.7$, adopting
a true distance modulus of 18.9 (van den Bergh 2000b) and typical $A_V=0.3$~mag extinction (Massey et al.\ 1995). 
This luminosity is just what is expected for  
for early-type WN stars (Vacca \& Torres-Dodgen 1990) in the LMC or Milky
Way, although such stars tend to have considerably stronger EWs (Conti \&
Massey 1989).  Indeed, the ave EW of He~II~$\lambda 4686$ of the LMC WNs
is $-100$~\AA, and we estimate that such a star would exhibit a magnitude
difference of -1.7~mag, and be detected at the $5\sigma$ level at $V=17$
($M_V=-2.2$).

The results of our analysis were encouraging.  With the exception of HD~5980
(which was saturated), all of the known SMC W-R
stars were found repeatedly on every field they appeared, typically with 
significance levels of 30$\sigma$ or more.  For instance, the strong-lined, bright W-R 
star AB4
($v=13.2$) was found with a {\it WN--CT} difference of $-$0.80~mag, at a
significance level of 78$\sigma$ and 87$\sigma$ on the two overlapping frames
on which it appeared. The faintest
previously known W-R star, the one discovered by Morgan et al.\ (1991),
with $v\sim 15.3$, was identified as a W-R ``candidate"
on all 3 fields in which it
appeared, with a {\it WN--CT} difference of $-0.42$~mag 
at a significance level of
11$\sigma$, 16$\sigma$, and 12$\sigma$.  Although the W-R star
HD~5980 itself was saturated on our frames, the nearby strong-lined O7~If+ star Sk~80 was detected (due to He~II~$\lambda4686$
emission) on all four overlapping frames that included that region, 
with a {\it WN--CT}=$-0.11$~mag at a significance level of 23--29$\sigma$.  Thus
we had some confidence that our detection technique would find any remaining
W-R stars even if they were significantly fainter and/or weaker-lined than
the known ones.

\subsection{Followup Spectroscopy}
Spectra of all the good candidates were obtained on the night of (UT) 2000
Oct 10 with the CTIO 1.5-m CCD spectrometer. A 527 line mm$^{-1}$ grating
(No.~16) was used in first order with no blocking filter to provide spectral
coverage 3990--5950~\AA\ with a spectral resolution of 4.2~\AA\ (1.5~\AA~pixel$^{-1}$).  A 3 arcsec slit was used in order to provide reasonable
spectrophotometry.  The known SMC W-Rs (with the exception of HD~5980) were
also observed, as well as the O7~If star Sk~80.  On subsequent nights
(2000 Oct 11-14) additional data were taken on the most interesting stars
(including HD~5980)
at higher resolution using the 831~line~mm$^{-1}$ (No.~47) in second order
with a CuSO$_4$ blocking filter.  This set-up provided a wavelength coverage
of 4090--4750~\AA,
with 1.4~\AA~resolution (0.5~\AA~pixel$^{-1}$) with a 1 arcsec
slit.  Our reductions followed the usual procedures, with the flat-fielding
accomplished by exposures of the ``punto blanco" (3-4 hrs in the case of the
higher dispersion grating).  For the spectral extractions, we used the optimized
extraction routines within IRAF, with care being take to select clean sky
on other side of the object.  In the case of nebulosity we performed several
extractions in an effort to minimize the under- or over-subtraction of nebular
emission.

Not unexpectedly, most
of our candidates turned out to be bogus for one reason or another,
usually related to the poor spatial resolution of the Schmidt images; e.g.,
two stars of very different colors were unresolved on our 
frames.  We also expect spurious detections (particularly those
found on a single frame) simply due to the number of 5$\sigma$ outliers
expected given the large number of stars photometered.  Many of our
W-R candidates proved to be K-type dwarfs, 
but which were detected at high significance level due to  the strong, broad MgH~$\lambda4780$ feature 
which depressed the {\it CT} magnitude. A total
of $\sim 25$ unknown sources were examined, of which roughly
a third proved of interest.

We discuss our more successful discoveries in the following section.

\section{Results}
\subsection{The Wolf-Rayet Content of the Small Magellanic Cloud}
Two of our high-significance candidates proved to be Wolf-Rayet stars of
WN3 type. We list their properties in Table~1, along with
those of the other SMC W-Rs\footnote{We refer to these two newly found W-Rs as SMC-WR10 and SMC-WR11, consistent with the IAU
nomenclature recommendations, and suggest that the previously known W-Rs
(often referred to as ``AB1" through ``AB8" and ``Morgan's star") be known as
SMC-WR1 through WR9.}.  In addition to the spectrophotometry from our
spectra, we also have Curtis Schmidt {\it BVR} images of our fields
obtained on 1999 Jan 8 as part of our search for red supergiants in
the SMC; we have used these data to provide $V$ and $B-V$ photometry for
our stars.  Although ``line-free" {\it v} and {\it b-v} values for the
W-R stars would be preferable (Smith 1968b), we have used our spectrophotometry to substantiate that the corrections are minor for these
relatively weak-lined W-Rs, and we use the broad-band colors to correct
for reddening until we can obtain better calibrated spectrophotometry.
(The exception is the newly found W-R SMC-WR10, for which nebulosity 
significantly affected the $B-V$ colors, and hence we have used the
spectrophotometry.)
We adopt a single $(B-V)_o=-0.32$ for all the W-R subclasses (see Fig.~4
of Pyper 1966). A value of $A_V=0.1$ corresponds
to a realistic minimum reddening, and $A_V=0.3$ is typical for early-type
stars in the SMC (Massey et al.\ 1995); the average $A_V$ of the W-Rs is
about 0.7~mag, suggesting we may be overestimating the reddening 
correction slightly by this method. 
A true distance modulus of 18.9 (van den Bergh 2000b) was adopted in
computing $M_V$.

The SMC W-Rs were throughly discussed by
Conti, Massey, \& Garmany (1989), who used 
SIT-Vidicon spectrophotometry along with moderate resolution photographic
spectra to provide  descriptions of each of the 8 SMC W-R stars known then.
(See also Moffat 1988.)
For consistency, we reobserved all of the SMC W-Rs along with our newly
found ones, and reclassified them {\it ab initio}.  We are in good agreement
with Conti et al.\ (1989) and Moffat (1988) except as noted below.
Since these data have somewhat higher resolution and better
signal-to-noise we briefly comment on each of the SMC W-Rs here. We illustrate
their spectra in Figs.~\ref{fig:wrs} and \ref{fig:newwrs}.

All but SMC-WR4 show absorption lines in their spectra, including our two
newly identified W-Rs.  Conti et al.\ (1989) emphasize that this could be
due  to these being binaries {\it or} due to their having thin stellar winds
allowing photometric absorption.  Moffat (1988) suggests that five of
these (WR3, WR5, WR6, WR7, and WR8) are double-lined spectroscopic binaries.
Of these, orbits have been determined for only three of these: WR6
(Hutchings et al.\ 1984),
WR7 (Moffat 1988; Niemela \& Morrell 1999), and WR8 (Moffat, Niemela, \& Marraco
1990),
with the issue for WR5 (HD~5980) very complicated (Koenigsberger et al.\ 2000
and references therein).
A  radial velocity study of the SMC W-Rs is currently underway, and for
now we will refer to the ones without established orbits as ``WR+abs" and
separately describe the absorption spectra.

{\bf SMC-WR1:} At $V=15.1$ this was the faintest of the original
eight SMC W-Rs.  The star has strong He~II~$\lambda4686$ and
N~V$\lambda\lambda4603,19$ emission, but no N~IV~$\lambda4058$ or N~III~$\lambda\lambda4634,42$ emission. (Conti et al.\ 1989 note the presence of N~IV~$\lambda3480$.) He~II emission at~$\lambda4859$ and~$\lambda5411$
are visible, and the W-R 
star is readily classified as ``WN3". A faint absorption
spectrum was reported by Moffat (1988), described as ``O4:".  On our
spectra we see only He~II absorption lines ($\lambda\lambda4200, 4542$)
plus the Balmer lines, and we call the absorption component O3-4.

{\bf SMC-WR2:} This star has narrow emission at He~II~$\lambda$ 4686,
N~IV~$\lambda4058$;  N~III~$\lambda\lambda4634, 42$ is weakly 
present in emission, and we classify the star as WN4.5.  
In addition emission at He~II
$\lambda4859$ and~$\lambda5411$ is present.  C~IV~$\lambda5808$ is
clearly present in our spectrum.  Its strength is too weak ($20\times$)
to consider this a transitional WN/C type
(see Fig.~5 of Conti \& Massey 1989) by 
Galactic or LMC standards, although its singular presence in this star
in the metal-poor SMC suggests that this may actually 
be ``on its way" from WN to WC type. Conti et al.\ (1989) note 
this star shows no
discernible radial velocity variations.  They 
describe the absorption
spectrum as being of ``early O type", in agreement with our spectrum,
which we tentatively describe as O3, based upon a high-dispersion,
high S/N 4-m spectrogram.  Walborn (1986) compares the spectrum to 
that of Sk$-67^\circ22$, both of which he classifies as O3If*/WN6.
We show such a comparison in Fig.~\ref{fig:sk22}, at considerably higher
resolution and signal-to-noise.  Although the similarity of the spectra
is striking, there are some notable differences, primarily that the emission
in SMC-WR2 is considerably stronger both at N~IV~$\lambda 4058$ and at 
He~II~$\lambda 4686$.  Consistent with this is the fact that HeII/H$\beta$
is in emission in SMC-WR2, but in absorption in the O3If*/WN6 star. 
We are unfamiliar with any published spectra of Sk$-67^\circ22$ in the
yellow-red, but would predict that He~II and C~IV would be found in
absorption, rather than in emission.  We propose that SMC-WR2 is
a {\it bona-fide} Wolf-Rayet star (e.g., an evolved object), 
while Sk$-67^\circ22$ is a very high luminosity O3 star with accordingly
very strong winds.  Detailed atmospheric modeling of the latter is currently
underway.

{\bf SMC-WR3:} At first glance, this star is readily classified as
``WN3+abs": there very strong He~II~$\lambda4686$ and N~V~$\lambda\lambda 4603,19$ emission lines, as well as the He~II~$\lambda4859$ and~$\lambda5411$
in emission.  We were puzzled by the very faint, but unarguable presence of
N~III~$\lambda\lambda4634,42$ emission in our spectrum, given the total
absence of N~IV~$\lambda4058$.  Such a possibility is not allowed for in
the classification scheme. Admittedly our signal-to-noise is higher than
that used to develop the classification criteria.
The faint absorption spectrum is
impossible to classify given emission at He~II~$\lambda4542$; however,
He~I~$\lambda4471$ is clearly present. Moffat (1988) argues that this
star is an SB2 with a period longer than a week, but an actual orbit has never
been published for this star.  We speculate that the very faint
N~III emission may belong
to the O-type star if the spectrum is truly composite.

{\bf SMC-WR4:} shows
moderately strong N~IV~$\lambda4058$ comparable to N~III~$\lambda4634,42$
emission, with weak N~V~$\lambda4603,19$ emission. Thus
we classify this star as WN6.
Conti et al (1989) described the star as ``WN4.5" (due
apparently to weak or absent N~III), while in our spectrum we see
N~III roughly equal in strength to N~IV.
Based upon photographic spectra obtained at about the same time,
Moffat (1988) had called the star a WN6, so we
believe this is not a real spectral change; this is also consistent
with the description given by Walborn (1977).
There are no absorption lines present other than P Cygni absorption components
to N~V.

{\bf SMC-WR5:} HD~5980 underwent an ``LBV-like" outburst (Barba et al.\ 1995), after which its spectrum was no longer ``OB+WN3". Niemela, Barba,
\& Morrell (1999) describe its spectrum as having evolved to a very
late WN type, which they call a ``WN11".
We have only
a higher resolution spectrum which does not include
coverage of the N~IV~$\lambda4058$ region, but both N~III and
N~V are present, and we might call this a WN5 without additional
information.  Strong lines of He~II
are present.
Our spectrum does indicate a a somewhat earlier 
type than the latest (1997) spectrum illustrated by Niemela et al.\ 
(1999).  Doubtless further monitoring will determine if the star
is evolving back towards earlier type or not.  The photometry we give
is from 1999 Jan; we lack photometry of this star
in its present state, but an intensive monitoring campaign by 
others (i.e., Niemela et al. 1999) is currently underway.

{\bf SMC-WR6:} Our spectra show He~II~$\lambda4686$ and modest
N~V~$\lambda\lambda4603,19$ emission, without either N~IV or N~III,
leading to the ``WN3" classification.
A strong absorption-line spectrum is present, which we classify as O7 type.
Moffat (1988) refers to the companion as O6.5, consistent with our designation.

{\bf SMC-WR7:} Conti et al.\ (1989) refer
to SMC-WR7 simply as ``WN+abs"; our high signal-to-noise
 spectrum shows no lines of
nitrogen, and we would call this a WN2+abs using the ``extended" criteria
of van der Hucht et al.\ (1981) that WN2 have weak or absent N~V.  We do
see He~II~$\lambda\lambda4542, 5400$ emission, consistent with the admonishment
that WN2 stars have ``strong He~II". A strong absorption spectrum is
present, which we classify as O6 type, similar to the ``O7:" description
of Moffat (1988).

{\bf SMC-WR8:} The only WC-type Wolf-Rayet known in the SMC, the
optical spectrum is dominated by C~III~$\lambda4650$ and C~IV~$\lambda5808$.  No C~III~$\lambda5696$ is seen, but strong O~V~$\lambda5592$ is present, and
would lead  to a WC4 classification.  Our spectrum does not go far enough
into the blue to detect the strong O~VI~$\lambda3820$ emission which led
to this being considered an WO4 star (Moffat, Breysacher, \& Seggewiss 1985),
but OVI $\lambda 5920$ is readily visible in the yellow-red, as expected
(Crowther, De Marco, \& Barlow 1998).
The absorption component was described by Moffat et al.\ (1985) as O4-type,
in agreement with the ``O4~V" classification by Massey et al.\ (2000).

We next consider the spectra of the Wolf-Rayet star found by 
Morgan et al.\ (1991) and our two newly found W-Rs.  The spectra are shown
in Fig.~\ref{fig:newwrs}.

{\bf SMC-WR9:} Morgan et al.\ (1991) classify 
SMC-WR9 as ``WN2.5+abs"; our spectrum of this star shows modest N~V emission
in good accord with their spectrum.  We classify the star as WN3+abs,
a difference due more to our interpretation of the spectral subclasses
than with any disagreement with the  description by Morgan et al.
We classify the absorption component as O3-4, based on strong He~II
absorption and an apparent lack of He~I~$\lambda4471$.  This is slightly
earlier than the ``O5:" designation of Morgan et al.

{\bf SMC-WR10:} This star is embedded in very strong nebulosity, the NW
knot of e12 (NGC~249), illustrated on plate 101V of Hodge \& Wright (1977).
This nebular
doubtless contributed to the difficulty of identifying the star as a W-R
via objective prism
techniques.  Our spectrum shows He~II~$\lambda4686$ with a similar
full-width-at-half-maximum (fwhm) and equivalent width to the other W-R WNs.
N~V$\lambda\lambda4603,19$ emission
is strongly present, as is He~II emission at~$\lambda4859$ and~$\lambda5411$.
There is no sign of N~IV~$\lambda4058$ or N~III~$\lambda4634, 42$, making
it of WN3 type.  Our higher resolution spectrum shows absorption at
He~II~$\lambda\lambda4200, 4542$, as well as H$\gamma$; we classify the
absorption component as O3.

{\bf SMC-WR11:} Our other newly found W-R star is fairly weak-lined,
and is the only star with appreciable reddening, but is otherwise 
unremarkable.  The star is not a member of any OB association or H~II region,
nor is it particularly crowded; only its faintness and weak lines can account
for its lack of previous attention.
The presence of strong N~V~$\lambda\lambda4609,20$
emission and the lack of either N~IV~$\lambda4058$ or N~III$\lambda\lambda4632,42$ argues again for the WN3
subclass; Balmer absorption and He~II~$\lambda4200, 4542$ is likewise
revealed on the high-resolution spectrum, leading us to classify the
absorption as O3-4 due to the apparent lack of He~I~$\lambda4471$.

We provide finding charts for all 11 of the SMC W-Rs in Fig.~\ref{fig:fcwrs}.

\subsection{Other Finds}

We list in Table~2 the other stars we found of interest, and discuss these
here.  Finding charts are given in Fig.~\ref{fig:fcothers}.

\subsubsection{Of-type Stars}

Wolf-Rayet stars are not the only early-type stars with He~II~$\lambda4686$
emission, of course. The most luminous O-type stars (e.g, luminosity class ``I")
also show He~II~$\lambda4686$ emission, as well as N~III~$\lambda\lambda4634,42$ emission. We aimed our selection to be sensitive to the weakest-lined W-Rs, and it is therefore not surprising that some strong-lined
Of-type stars were detected. Two of these were previously known:
AV~220, 
of type O7~If (Garmany, Conti, \& Massey 1987) recently reclassified
as ``O6.5f?p" by Walborn et al.\ (2000), and Sk~80, a
O7~If+ star with strong He~II~$\lambda 4686$ emission
(Walborn \& Fitzpatrick 1990). However, our survey also
uncovered four previously unknown Of stars, and we list these in Table~2.
One of these is of type O5f?p, making it one of the earliest-type stars
known in the SMC.  We have adopted $(B-V)_o=-0.30$ in determining the
correction for interstellar extinction.

Interestingly three of the four newly found Of stars do not appear
in the Azzopardi \& Vigneau (1982) catalog
of SMC members.  At $V\sim 13.4-14.0$ these are considerably brighter than the
$V\sim15$ plate limit of the Azzopardi \& Vigneau (1975, 1982) survey; all
three are within the survey region (two on their chart 5, and one on their
chart 4).  All three are located in the bar, arguing that perhaps
crowding may have
made their detection difficult.

Here we discuss the four
stars in turn.

{\bf MA93-344:}  This Of star was cataloged as an H$\alpha$ emission-line star
by Meyssonnier \& Azzopardi (1993).  We
show the spectrum in Fig.~\ref{fig:ofs}.  Our efforts to obtain a suitable
high-resolution spectrum of this star failed due to clouds on our last 
observing night; the lower resolution spectrum we show in Fig.~\ref{fig:ofs}
reveals a late-O spectrum with broad He~II~$\lambda 4686$ emission with
an absorption feature superposed.  There is no trace of N~III~$\lambda 4634,42$
emission. Were this star a Galactic star we would be tempted to cite a binary
explanation, with the companion possibly of W-R type.  Si~IV~$\lambda 4089$
absorption
is strong, supporting the interpretation that the O star is of luminosity
class ``I".  The absorption spectrum is consistent with an O9 or O9.5-type, although
in the LMC and Milky Way He~II$\lambda$ emission is not see in types later
than O8.5. (We are indebted to Nolan Walborn for this reminder.)
Keeping in mind that we do not fully understand the behavior
of ``Of" features in the low metallicty environment of the SMC, we tenatively 
describe the star as an O9~If, but do not exclude the possibility this
star is a binary with a twelfth W-R companion.  A better spectrum is needed
to resolve this issue.


{\bf Anon-1:} This star has {\it very} strong emission for an Of-type star;
we classify this star as O5~f?p, after the small class of 
such peculiar objects discussed
by Walborn (1973).  We recall that one of the two known Of stars
found in our survey is AV~220, which Walborn et al.\ (2000) also put in
this class; only five such objects are known anywhere.
Such stars have broad absorption,
with narrow emission at the Balmer lines in addition to strong, narrow He~II~$\lambda 4686$ emission, with the ``?" intended to express doubt
that the He~II emission is due to high luminosity. 
In Fig.~\ref{fig:anon1} we compare the spectrum
to that of the O7~If+ star Sk~80. We obtained
several spectra of Anon-1, and it does not {\it appear} to be a binary,
in that the spectral morphology did not change over several nights.
The very narrow He~II~$\lambda 4686$ line (6~\AA~fwhm) argues that this is
an Of rather than a composite W-R type.  The $-5$~\AA~equivalent
width is strong for a Of star, but not unprecedented; the Galactic O8~If star HD~152408 does have
comparable He~II $\lambda$ line strength (Conti \& Leep 1974).  Walborn
et al.\ (2000) cites HD~108 and HD~148937 as the prototypes of the ``Of?p"
class; the equivalent width of He~II~$\lambda 4686$ in our SMC object
is 4$\times$ stronger than that in HD~108
(Conti \& Leep 1974); similar quantative information on HD~148937 is
lacking.
Examination of the UV stellar wind lines in other ``Of?p"
suggest that such objects are not actually supergiants (Walborn et al.\ 2000);
such a conclusion is in accord with the 
the modest absolute magnitude of this object  ($M_V=-5.3$).  The emission
is presumably due to a shell.  The star is located in OB association 35 (Hodge 1985), along with S18, the
LBV we will discuss in the following section.  

{\bf Anon2:} The spectrum is that of an O7~If. N~III$\lambda\lambda 4634,42$
is anomalously strong, and He~II $\lambda 4686$ shows absorption superposed
on emission.  Such a star would be called an ``Onfp" in Walborn's
nomenclature (Walborn 1973), or Oef in Conti's (Conti \& Leep 1974).


{\bf AV398:} The fourth newly found Of star, AV~398, has been classified as ``O9Ia:" from
an {\it IUE} spectrum by Neubig \& Bruhweiler (1997), who quote the optical
spectral type as ``B2" based upon the objective prism classification of
Azzopardi \& Vigneau (1975). Our spectrum clearly reveals a normal
O8.5~If star (Fig.~\ref{fig:ofs}). Interestingly this star has long been used
as a ``typical bar line of sight" star for studies involving the extinction
law of the SMC (most recently Zubko 1999).  Such studies eschew
emission-line stars (see discussion in Rodrigues et al.\ 1997); apparently
the objective prism ``B2" classification was taken literally.

\subsubsection{Two Other Emission-line Stars}

Our survey detected the star S18=AV~154, a B[e] star that has many
LBV-like characteristics (Morris et al.\ 1996). 
S18 was detected on three
overlapping fields, observed during three
consecutive nights (UT 1999 Oct 25-27), and during this
time the {\it CT} magnitude faded from 13.59 to 13.74. Our spectrum was
taken a year later.
At the first look at the spectrum we were confused that the star had been
a candidate at all, as there was no detectable He~II~$\lambda4686$ present,
despite numerous other emission lines.
(The object was a 95$\sigma$ candidate!)
However, the spectrum shown by Shore, Sanduleak, \& Allen (1987) clearly does
show strong He~II $\lambda 4686$.  Thus we conclude that He~II $\lambda 4686$
emission was strongly present when our imaging was taken in 1999 Oct, despite
its absence a year later when we obtained the spectrum.
Its absence during 2000 Oct is not
without precedent, however, 
as Zickgraf et al.\ (1989) also reported the
line as having been missing in a 1987 spectrogram. This star is discussed
extensively by Morris et al.\ (1996), who emphasize the need for further
monitoring.  Here we draw attention to its spectral similarity to
S~Dor, the archetypical LBV, during its 1996 Oct ``low" state (Massey
2000).  We compare the two in Fig.~\ref{fig:s18}.   Given the photometric
variability, high luminosity, spectral changes, and its spectral
similarity 
spectrally to S~Dor, one would be hard-pressed not to consider this an LBV.

The star is located in OB association 35 (Hodge 1985), the same association
which our newly discovered O5f?p star is found.  Our study of the
environments of LBVs in the Magellanic Clouds and Milky Way suggest that
the LBVs which are members of coeval regions are nearly always found with
extremely massive stars.  A detailed study of h35 is doubtless warranted.

The last star of interest was included in the list of H$\alpha$ emission-line
stars given by Meyssonnier \& Azzopardi (1993), their number 1748.  Nothing
else was previously known about the star.  Its spectrum, shown in Fig~\ref{fig:symbiotic}, was clearly of a symbiotic star.  It does not appear in the
recent catalog of symbiotic stars by Belczynski et al. (2000), and appears to
be only the seventh such object known in the SMC. 
Howard Bond and Tony Keyes
kindly pointed out its similarity to that of the Galactic symbiotic
star AG~Dra, illustrated in Fig~A.14 of Kenyon (1986).
With $V=16.61$, we would estimate an absolute magnitude
of $-2.6$, not inconsistent with the typical $M_V=-2$ quoted by Kenyon (1986).

\section{Discussion}

Our survey discovered two previously unknown Wolf-Rayet stars, bringing the
total number in the SMC to 11.  Although we cannot preclude a Wolf-Rayet
star or two as having been overlooked in our survey, particularly in crowded
regions, we believe this study underscores that there is {\it not} a significant
number of W-Rs waiting to be discovered in the SMC.
Thus W-R stars are under-abundant in the SMC by a
factor of 3 (per unit luminosity)
compared with the LMC. 
As we argue in Section 1, we believe that this supports the view that
it requires a higher mass to become a W-R in the SMC than it does in the LMC,
due to the effect of lower metallicity on stellar winds.

It is iteresting that both of the newly-found W-Rs show the presence of
absorption lines.  
When Breysacher \& Azzopardi (1979) first announced the discovery of four
additional W-R stars in the SMC, and noted that all appeared to show some
sign of OB companions, the inital reaction was that probably Roche-lobe
enhanced mass-loss was crucial to a massive star becoming a W-R at low
metallicity---that presumably stellar winds by themselves were not 
sufficient.  Subsequently, Conti et al.\ (1989) re-examined this point,
noting that the case for {\it bona-fide} binarity was poorly established
for most of the SMC W-Rs.  An alterantive explanation for the prevelance
of absorption lines in the spectra of the SMC W-Rs could simply be that
the stellar winds are weaker, allowing normal photospheric lines to be seen.
Over a decade later, orbit solutions still exist for only the same three
SMC W-Rs, with the situation for a fourth (HD~5980)
made even more complex by its LBV outburst.  
A radial velocity study is currently
underway for all of the SMC W-Rs, which we expect to be complete after
the 2001 Magellanic Cloud observing season.   We do note that the statistics
for the SMC are very different than for the LMC: for the latter, an
absorption-line spectrum is seen only for the brighter systems, consistent
with the fact that an OB companion will make the system brighter than
on average (Vanbeveren \& Conti 1980).  The presence of absorption lines
in even the faintest of the SMC W-Rs {\it could} suggest that they are in
fact all binaries---or equally well that the absorption lines are seen thanks
to a weak stellar wind.

We note that the SMC W-Rs are, on average, more luminous than their 
counterparts
in the Milky Way or LMC.  Ignoring the ``+abs" systems, Vacca \& Torres-Dogden
(1990) find and average $M_v=-3.7$ for the early-type WNs; the average for
the same type stars in the SMC is $-5.3$, where we of course have to include
``+abs" systems as most of the stars show absorption lines.  Such an average
is considerably more luminous even than the $-4.6$ adopted by Conti \& Vacca
(1990) for the ``WNE+abs" systems.  This cannot be a selection effect, as
we argued earlier, as our survey went penty deep enough to find both
weaker-lined, fainter stars.  This {\it could} mean that in general the
W-Rs in the SMC do have binary companions which dominate the continuum. Or,
it could simply be a reflection of the higher luminosity requirement needed
for stellar winds to lead to the W-R stage at lower metallicities.

Our discovery of an O5f?p star, 
as well as three other hitherto unknown Of stars,
emphasizes that much spectroscopic work remains to be done in the Magellanic
Clouds.

\acknowledgements
It is a pleasure to thank the CTIO support staff for their help. 
The Curtis Schmidt
imaging  kindly came about via director's discretionary time.
A.S.D. participated in these studies partially through 
the Research Experiences for Undergraduates program, which was supported by
the National science Foundation under grant 9988007 to Northern Arizona
University. Drs. Peter Conti, Deidre Hunter, Nichole King, and Nolan
Walborn all made thoughtful and useful comments on the manuscript, as did
an anonymous referee.

\clearpage
\begin{deluxetable}{l l c c c c c c c c c c l}
\renewcommand{\arraystretch}{0.6}
\tabletypesize{\scriptsize}
\rotate
\tablewidth{0pc}
\tablenum{1}
\tablecolumns{13}
\tablecaption{Wolf-Rayet Stars in the SMC}
\tablehead{
\colhead{Star}
&\colhead{Other ID}
&\colhead{$\alpha_{2000}$}
&\colhead{$\delta_{2000}$}
&\colhead{OB/HII\tablenotemark{a}}
&\colhead{Spectral Type}
&\colhead{Abs.}
&\colhead{$V$}
&\colhead{$B-V$}
&\colhead{$M_V$}
&\multicolumn{2}{c}{He~II or C~III\tablenotemark{b}} 
&\colhead{Comment}  \\ \cline{11-12}
&&&&&&&&&&\colhead{EW (\AA)}
&\colhead{FWHM(\AA)} \\
}
\startdata
SMC-WR1 & AV~2a & 00 43 42.23 &  $-$73 28 54.9 & no & WN3+abs  &O3-4 & 15.14 & $-0.04$   &$-4.6$  &  $-$28 & 21 & Weak abs. \\
SMC-WR2 & AV~39a& 00 48 30.81 &  $-$73 15 45.1 & near h15 &WN4.5+abs&O3 & 14.23 & $-$0.15   &$-5.2$  &  $-$15 & 12\\
SMC-WR3 & AV~60a& 00 49 59.33 &  $-$73 22 13.6 & near h17 &WN3+abs &\nodata&14.48 & $-$0.10  &$-5.1$  &  $-$53 & 26 & Very weak abs. \\
SMC-WR4 & AV~81, Sk~41& 00 50 43.41 &  $-$73 27 05.1 &h21&  WN6p&\nodata &13.35& $-$0.16&$-6.2$ & $-$45 & 15 & N~V abs (P Cyg)\\
SMC-WR5 & HD~5980 & 00 59 26.60 &  $-$72 09 53.5&NGC~346=h45 & WN5&\nodata & 11.08& +0.03 & $-$8.9 &  $-$85 & 18 \\
SMC-WR6 & AV 332, Sk 108 & 01 03 25.20 &  $-$72 06 43.6 &h53 & WN3+abs &O7& 12.30 &$-$0.15 & $-7.1$&  $-$8 & 28 \\
SMC-WR7 & AV~336a & 01 03 35.94 &   $-$72 03 21.5 & h53&  WN2+abs & O6&12.93&$-$0.05 &$-6.8$ &  $-$16 & 27 \\
SMC-WR8 & Sk 188 & 01 31 04.22 &  $-$73 25 03.9 & NGC~602c=h69 & WO4+abs &O4~V& 12.81 &$-$0.14 &$-6.6$ &  $-$76 & 71 \\
SMC-WR9 & Morgan et al.\ & 00 54 32.17 &  $-$72 44 35.6 & no & WN3+abs &O3-4& 15.23 &$-$0.13&$-4.3$ &$-22$ & 24\\
SMC-WR10 &  & 00 45 28.78 &  $-$73 04 45.2 & NGC~249(e12)&WN3+abs & O3-4&15.76:&$-$0.08: &$-3.6$ &  $-$24 & 23 & Strong neb.\\
SMC-WR11 &  & 00 52 07.36 &  $-$72 35 37.4 & no & WN3+abs & O3-4&14.97&+0.18&$-5.5$ &  $-$14 & 25 & \\

\enddata
\tablenotetext{a}{OB associations designation (h) are from Hodge (1985), and
emission-line regions (e) are from the Hodge \& Wright (1977) atlas.}
\tablenotetext{b}{The equivalent width (EW) and full-width-at-half-maximum
(FWHM)
of He~II~$\lambda4686$ is given for
the WNs; that of C~III~$\lambda4650$ is given for the WO4 star.}
\end{deluxetable}

\clearpage
\begin{deluxetable}{l l c c c c c c l}
\renewcommand{\arraystretch}{0.6}
\tabletypesize{\footnotesize}
\tablewidth{0pc}
\tablenum{2}
\tablecolumns{9}
\tablecaption{Non W-R Stars of Interest}
\tablehead{
\colhead{Star}
&\colhead{Other ID}
&\colhead{$\alpha_{2000}$}
&\colhead{$\delta_{2000}$}
&\colhead{OB/HII\tablenotemark{a}}
&\colhead{$V$}
&\colhead{$B-V$}
&\colhead{$M_V$}
&\colhead{Spectral Type}
}
\startdata
\sidehead{Of-type}
MA93-344 & Lin 146 &00 50 11.30& $-$73 20 54.2&  no & 13.83& $-$0.2:& $-$5.4& O9~If  \\ 
Anon-1  & \nodata & 00 53 29.95& $-$72 41 44.2& h35 & 13.98& $-$0.18& $-$5.3 &O5f?p  \\ 
Anon-2 & \nodata  & 00 55 52.85& $-$73 22 35.3& no &13.41& $-$0.15& $-$6.0
&O7~If \\ 
AV398  & \nodata  & 01 06 09.78& $-$71 56 00.3& no &13.88& +0.09 & $-$6.2 &O8.5~If \\ 
\sidehead{Other}
S18 & AV 154 &00 54 09.56& $-$72 41 43.2& h35 &13.7var& +0.3& $-$6.8: &B[e] \\ 
MA93-1748 & \nodata &01 11 37.45& $-$71 59 01.8 & no &16.61 & +1.04 &$-$2.6&  Symbiotic \\ 
\enddata
\tablenotetext{a}{OB associations designation (h) are from Hodge (1985).} 
\end{deluxetable}

\clearpage

\figcaption{The region surveyed for Wolf-Rayet stars is shown by the 12 
overlapping squares, each of which is 1.3$^\circ$ on a side. The area surveyed by objective prism by
Azzopardi \& Breysacher (1979) is shown
by the 5 circles.  The locations of the 11 W-R stars are indicated.  The
diagonal streak is an artifact on the Digitized Sky Survey image used to
construct this diagram.
\label{fig:smc}
}

\figcaption{The normalized 
spectra of the 
8 Azzopardi \& Breysacher (1979) W-R stars are shown,  with the principal
emission lines indicated.  In (a) we show the weaker-lined stars, and in
(b) we show the stronger-lined stars; the scaling is compressed
a factor of 3 compared to that of (a).
\label{fig:wrs}
}

\figcaption{The spectrum of SMC-WR2 is compared to that of the LMC star Sk$-67^\circ22$,
which is classified as O3If*/WN6-A by Walborn (1986).  The spectra were
obtained with the CTIO 4-m on 2000 Jan 4 and 5, respectively. \label{fig:sk22}}

\figcaption{The normalized spectra of our two newly-found SMC W-R stars
WR10 and WR11 are compared to that of the star found by Morgan et al.\
(1991), WR9.  The scaling is the same as in 2(a).  \label{fig:newwrs} }

\figcaption{Finding charts are presented for all 11 of the SMC W-Rs. Larger
scale versions for SMC-WR1 through 8 appeared in Azzopardi \& Breysacher
(1979), and for WR9 in Morgan et al.\ (1991).  The ones shown here are
5~arcmin on a side, and were made from the Digitized Sky Survey; the circles
marking the W-Rs are 20 arcsec in diameter. The following should be noted:
WR4 is the brightest star in a small group.
WR6 is slightly crowded; the correct star is the brighter star to the NW.
WR7 has a faint companion to the NE.
WR8 is the brightest star in the cluster NGC 602c, and a larger-scale
chart can be found in Fig.~4 of Westerlund (1964); it is the star labeled 17.
\label{fig:fcwrs}
}
\figcaption{Finding charts for the other interesting objects we found are
given here.  The finding charts are 5 arcmin on a side, and the circle marking
the object is 20 arcsec in diameter.
\label{fig:fcothers}
}

\figcaption{The spectra of three of our newly found Of stars are shown; the
fourth is shown in the following figure. 
\label{fig:ofs}
}

\figcaption{The spectrum of the newly found O5f?p is compared to the
extreme O7~If+ star Sk~80.  The latter spectrum was taken with our lower
dispersion grating.
\label{fig:anon1}
}

\figcaption{The spectrum  of the B[e] star S18 obtained in 2000 Oct
is compared to that of S~Dor when it was in its ``low state"
a 1996 Oct.  The later spectrum comes from Massey (2000).
\label{fig:s18}
}

\figcaption{The spectrum of our newly found symbiotic star is shown.
We show the spectrum preserving the flux scale (units are ergs~s$^{-1}$~Hz$^{-1}$).  The absolute flux has been adjusted by a factor
of 1.5$\times$ to account for slit losses and bring the photometry into
agreement with the {\it CT} magnitude.
The spectrum closely resembles that of AG~Dra; see Kenyon (1986), Fig.~A14.
The star was noticed as an H$\alpha$ emission-line object by
Meyssonnier \& Azzopardi (1993), who catalog it as number 1748.
\label{fig:symbiotic}
}

\end{document}